# Narratives in economics

Michael Roos, Ruhr-University Bochum

Matthias Reccius, Ruhr-University Bochum

October 26, 2022

___

**Abstract**

There is growing awareness within the economics profession of the important role narratives play in the economy. Even though empirical approaches that try to quantify economic narratives are getting increasingly popular, there is no theory or even a universally accepted definition of economic narratives underlying this research. First, we review and categorize the economic literature concerned with narratives and work out the different paradigms at play. Only a subset of the literature considers narratives to be active drivers of economic activity. To solidify the foundation of narrative economics, we propose a definition of *collective economic narratives,* isolating five important characteristics. We argue that, for a narrative to be economically relevant, it must be a sense-making story that emerges in a social context and suggests action to a social group. We also systematize how a collective economic narrative differs from a topic and from other kinds of narratives that are likely to have less impact on the economy. With regard to the popular use of topic modeling, we suggest that the complementary use of other methods from the NLP-toolkit and the development of new methods is inevitable to go beyond identifying topics and move towards true *empirical narrative economics*.

___

# 1 Introduction

Sacco (2020) asks whether we are observing a `narrative turn´ in economics and answers that economics is probably not ready for a narrative turn yet. Nevertheless, he sees great potential for economics in the analysis of narratives, because they are important determinants of human behavior. A narrative turn took place in several social sciences as part of the post-positivist movement, which questioned and challenged the positivist approach to the study of social phenomena (Goodson and Gill, 2011). Narrative approach means that researchers became interested in subjective human understanding and sense-making, but also in social discourse. Narrative analysis entered the social sciences in the 1980s and 1990s, among them political science, psychology, sociology and science studies (Czarniawska, 2004). At the same time, Deidre McCloskey was a pioneer with her analysis of rhetoric and storytelling in economics (McCloskey, 1985; 1990a; 1990b; 1994). McCloskey argues that against the official methodological doctrine of modernism, economists use metaphors and stories to persuade other economists. Every mathematical model requires a story that connects the abstract equations to reality. She stresses how the language of economic theory itself largely consists of metaphors from the natural sciences – think of *velocity* of money, *elasticity* of demand – which is a practice seldomly reflected by economists. Rhetorical devices, according to McCloskey, do not merely dress up economic thought, but often yield its very substance, like in the case of Gary Becker´s description of children as durable goods. (McCloskey, 2002). She claims that economists are poets and novelists and suggests that literary thinking might improve applied economics (McCloskey 1985). However, it was not until Robert Shiller's (2017) Presidential Address at the American Economic Association that general attention to narratives and *narrative economics* (Shiller, 2019) rose in the profession. In contrast to McCloskey, who emphasizes the role of storytelling for the way economics is done, Shiller argues that narratives are important, because they help explain economic phenomena such as recessions or bubbles in financial markets. While narrative economics is a promising endeavor, it is not an easy field for newcomers. As Sacco (2020) rightly argues, there are still many loose ends and there is great benefit from interdisciplinary work. Such interdisciplinary exchange with researchers from other social sciences or even from the humanities is not easy at all due to different concepts of narratives. Even within economics there are a variety of uses of the term *narrative* and of claims of what we can explain with the concept.

With this paper, we want to generate more conceptual clarity in narrative economics. We review the strands of the economics literature that use the term narrative and show that the concept is not precisely defined. We propose a definition of *collective economic narratives* which we believe to be useful for research in economics and show how it can be applied. We link our concept to a large literature from literary studies, psychology, cognitive science, but also to older strands of institutional economics and political economy that are relevant for the understanding of narratives, but ignored by the more recent literature. As a final contribution, we discuss the tension between a clear theoretical definition and the difficulties of its empirical implementation. We argue that empirical research should be guided by precise theoretical concepts, despite problems of measurability and data availability.

The main claim of our paper is that the term *narrative* is not well-defined in the economics literature. For most authors a narrative is some kind of *story*, but it sometimes also has the meaning of *topic*. As we will show, empirical papers that try to measure narratives often simply identify topics. The precise use of terms is a prerequisite for scientific progress. When a research field is young, there might be benefits if terms and concepts are open to interpretation, because this can stimulate fruitful discussion and creativity. But after an exploratory stage, rigor is needed in order to determine what knowledge has been created and where the gaps are that should be filled by further research.

A scientific definition must be relevant for a specific community and a specific purpose. As said before, narratives are objects of investigation in many scientific disciplines, all of which have different aims of inquiry. For instance, linguists want to understand and consider narratives as an object of language. Economists, in contrast, are not interested in language per se, but only with regard to its relation to their main objects of inquiry such as the economic activities of production and consumption. Hence an economic definition of narrative can and should be different from a linguistic one. However, a definition should not contradict definitions in other fields. While different disciplines can use different definitions, it is not conducive to interdisciplinary exchange if the disciplinary definitions have nothing in common or even contradict each other. At least the main elements of a definition should overlap and should be used with a similar meaning. In Section 3, we propose a definition of *collective economic narratives* which is specific to the interests of economists, but also aligns well with how other disciplines think about narratives.

To illustrate the state of the art in narrative economics, it is informative to look at Shiller (2020), that defines economic narratives as "stories that offer interpretations of economic events, or morals, of hints of theories about the economy" (Shiller, 2020, p. 792). According to this definition, the main definiens is the term *story*, which itself remains undefined. Not every story is a narrative. Some stories "offer interpretations of economics events", for which Shiller uses the term *moral*. Alternatively, the story can offer "hints of theories about the economy". This definition is suggestive, but not precise, and hence itself requires interpretation. The "moral" element of the narrative definition might mean that narratives have an evaluative dimension, possibly linked to a certain suggested behavior. The "hints of theories" indicates that narratives also explain and contain statements about causal relations. Shiller (2020) proposes key words or phrases to measure six narratives which he believes to be associated with the U.S. macroeconomic evolution in the past 30 years: "Great Depression", "Secular Stagnation", "Sustainability", "Housing Bubble", "Strong Economy" and "Save More". He shows how the percentage of newspaper articles that contain these markers of narratives evolved since 1990 and gives a brief account of how he understands the narratives and their potential connection to the macroeconomy. While Shiller's ideas are suggestive and stimulating, they are also rather vague and subjective. For instance, on the alleged sustainability narrative, he writes: "This word [sustainability], as applied to conservation and climate change, went viral slowly over decades, from very small beginnings in the 1980s. It represents the idealism of the new generation, and logically leads to less intense spending" (Shiller, 2020, p. 797). This claim neglects that there are different conceptions of sustainability that by no means necessarily imply less spending. As Levy and Spicer (2013) argue the idea of sustainable lifestyles in the post-growth or degrowth sense with less consumer spending never took hold in the U.S.. Instead, sustainability conceptions related to green growth are much more popular, because investment into green technologies is seen as source of economic growth and new jobs.

We postulate that narratives are never independent from the people who invent and circulate them. If this claim is true, it poses a great challenge for the economic analysis of narratives, because we cannot simply look at words or texts alone in order to identify narratives. We also have to identify agents' *belief systems*, which give narratives their meaning.

In Section 2, we give an overview of how authors in economics use the term "narrative" in their research, before we propose our concept of collective economic narratives in Section 3. For our concept, we draw upon a rich economic literature that significantly predates Shiller's work, but is not mentioned by him. In particular, we find literature from political economy and institutional economics very helpful. Furthermore, we include literature from other disciplines such as literary studies, cognitive science and social psychology. In Section 4, we discuss two examples from the literature of how others use the term narrative and argue that these examples do not satisfy our definition. Section 5 discusses some challenges for quantitative empirical research that follow from our concept. and some technical approaches by which those challenges could be overcome. We present some wider implications of narratives in economics in Section 6. There are intradisciplinary implications for economics itself, which might follow McCloskey's advice to reflect more on its own use of narratives as a means of rhetoric and fiction. But there are also implications for how narrative economics could engage more with other disciplines. Section 7 concludes the paper.

## 2 Review of the economic literature using the term narrative

In this section, we provide review of the economic literature that uses the term "narrative". The purpose of this section is to show that there are a variety of concepts of narratives and of contexts in which the term is used. Note that we do not intend to review all strands of literature that are somehow related to the concept of narratives. Such literature will be discussed when we propose our own definition of narrative. Our main conclusion is that currently, we cannot speak of *the* narrative approach to economics or a coherent field of narrative economics. In fact, what we find is that there are different strands of the literature in which the term *narrative* is used in quite different ways. In order to support this claim, it is sufficient to present a selection of typical papers.

We searched for papers with "narrative*" in the title in the *economics* category of the Web of Science database and found 327 papers in the period from 1997 until May 2021. As shown in Figure 1, on average about 5 papers per year were published from the end of the 1990s until 2012. In 2013 a remarkable jump to 19 publications occurred and the number rose to 35 in 2020. Searching for "narrative*" in the abstract instead of the title shows a similar trend although the numbers are, of course, higher (see Figure 2). In the rest of the paper, we focus on the publications with "narrative*" in the title, because this is a harder criterion than when the term appears in the abstract.

Based on our reading of the abstracts we assigned the identified papers to seven categories as shown in Figure 2. As every categorization, this exercise can be debated. In several cases, papers could also have been assigned to a different category. This does not pose a problem to our argument since our categorization only serves as a heuristic to get an overview about the various meanings of narratives in economics so far.

**Figure 1: Count of papers with "narrative*" in title/abstract in Web of science category economics**

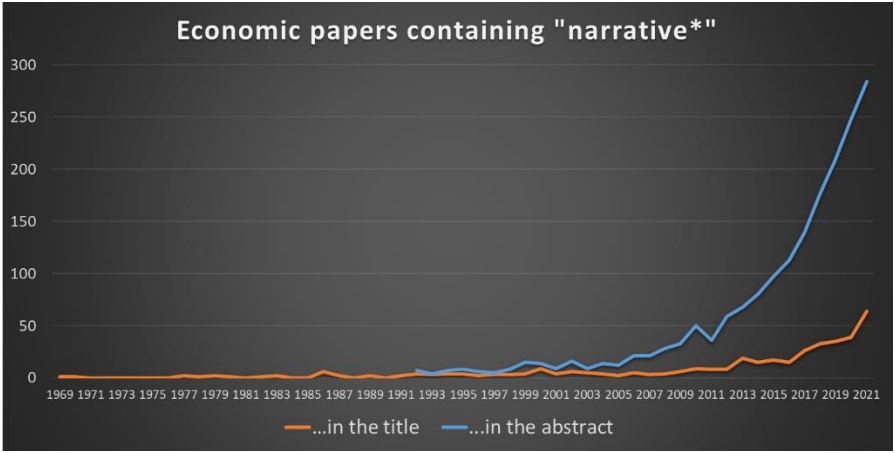

**Figure 2: Categorization of papers with "narrative*" in title in Web of science category economics**

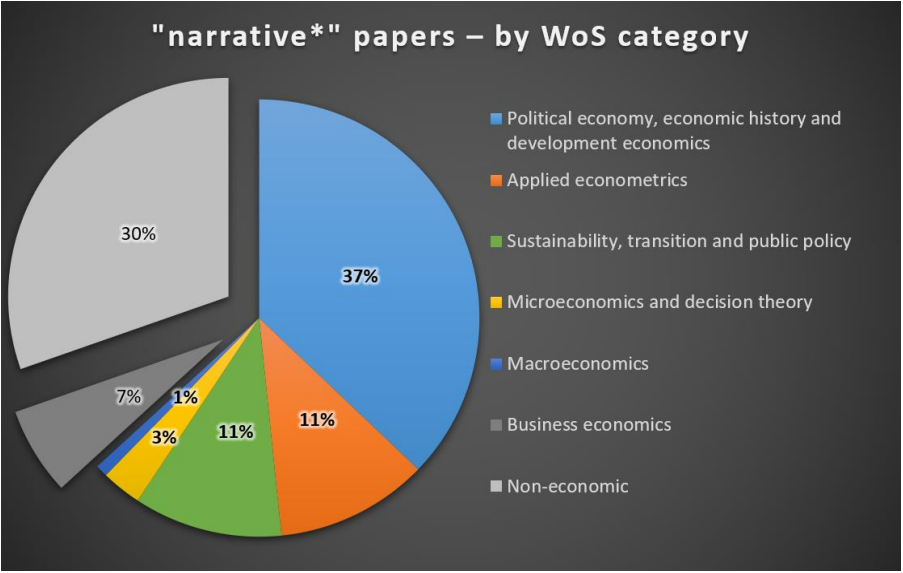

We classify a rather large fraction of 30 percent of all identified publications as non-economic although they are listed in the Web of Science category *economics*. This category also includes book reviews and some health and medical science publications that do not qualify as health economics. Another large share of the papers in this category belong to political science rather than economics. Because of its differing subject matter, the research pertaining to business economics has also been left out of the review.

In the following subsections we discuss what we consider to be the main ideas and uses of "narrative" in the categories (1) political economy, economic history and development economics, (2) macroeconomics, (3) applied econometrics (4) sustainability, transition and future thinking, and (5) microeconomics and decision theory.

## 2.1 Political economy, economic history and development economics

As mentioned in the Introduction, a narrative turn took place in political science already in the 1980s. As a result, it is not surprising that a large share of the identified papers falls within the broad scope of political economy. However, the lines between political economy and political science proper are blurry. In political science, narratives are assumed to be actively and socially constructed by policy stakeholders to serve a specific purpose (Miedzinski, 2018). They result from blending empirical facts with normative evaluations and goals (Majone, 1989).

The early political economy literature uses the term *narrative* not as an analytical category, but as a shorthand for a specific way in which a phenomenon is explained and interpreted, either in the research literature or by "conventional wisdom". In most of these cases, a dominant explanation of a particular issue is laid out, disputed and finally appended by a marginalized narrative whose importance is perceived to be underappreciated in the literature. As a result of this method, the status of the dominant narrative is reduced to a mere perspective rather than a proven fact. A good example of this use of *narrative* is Gottfried and Hayashi-Kato (1998) who deconstruct the success story of the Japanese economy in the post WWII era: While the dominant narrative stresses distinguishing labor market features such as lifetime employment guarantees as the most important factors of Japanese growth, Gottfried and Hayashi-Kato (1998) contest this conventional view by highlighting the important role of non-standard employment of women. Other similar examples are McMaster's analyses of "quasi-market" narratives in the context of welfare-state reforms and commodity narratives in health economics (McMaster, 2002; 2013), Good's take on economic performance in 19$^{th}$ and 20$^{th}$ century Europe (Good, 2002) and Hartmann's exposition on Neo-Malthusian scarcity narratives (Hartmann, 2010). This usage of the *narrative* concept is also present in a rather political strand of the development economics literature, where such dominant narratives often concern issues surrounding agriculture and land-use (Bergius et al., 2020; Dercon, 2013; Ellis and Manda, 2012; Fairhead and Leach, 1995) or the effects of trade and development aid (Engström and Hajdu, 2019; Gautam, 2019; Béné et al., 2010). Occasionally, the dominant narrative to be replaced is explicitly linked to serving political interests and the preservation of power (Mehta, 2001).

A few publications from this category explicitly claim to develop narratives themselves. Hoaas (1993) delivers a reflexive analysis of economic teaching principles that he refers to as a historical narrative of methodological changes contained in economic textbooks. Interestingly, the term *narrative* only appears once in the title of the paper. The goal in utilizing the term is simply to describe Hoaas´ own process of structuring the historical evolution in the teaching of economics. However, this paper and particularly some follow-up research (Hoaas and Madigan, 1999) also features what must be considered narrative summaries of central ideas and theories developed by important scholars of economics throughout history, starting with Smith´s famous "invisible hand". Roe (1995) works out two dysfunctional narratives in the context of African economic development and calls for the development of policy-relevant counter-narratives, including some of her own suggestions. To rise above the "Neo-Malthusian Doomsday Scenario"-narrative about population growth as the root of African stagnation, Roe (1995) suggests a counter-narrative prominently featuring the poor state of educational systems in the region and views experts in government and international organizations as responsible for inducing the needed reversal in thought patterns.

More recently, there was a shift in how *narrative* has been used in the political economy literature. Instead of standing for an interpretative story about the world, it denominates the stories that political agents in the socio-economic system use themselves. The shift, hence, is changing the scope of narratives from observers' interpretations of real-world phenomena to narratives as real-world phenomena themselves that are to be explained in scientific theories or formal models. Eliaz and Spiegler (2020) view political disagreements as a "clash of narratives''. They admit that the term *narrative* is vague and they provide a specific definition in the context of a formal network model. According to this definition, a narrative is a causal model that maps (political) actions to consequences, which can be represented by a directed acyclic graph. An example of such a directed acyclic graph is:

*trade policy -> imports from China -> employment*

which links U.S. employment to U.S. trade policy via its effect on imports from China in a causal story. In political debates different narratives are used that employ different intermediate variables and arrange variables differently in the causal scheme. Politicians offer competing narratives to the public, which selects certain narratives in order to form beliefs that are used to evaluate policies. The model describes an equilibrium process of narrative production and selection.

Antoci et al. (2020) are interested in how social influencers can affect the public opinion with competing narratives. They define narratives "at a very abstract level, … [they] might reflect a variety of possibilities: from purely fictional accounts that for some reason have become salient to the public opinion, to narratively biased accounts of real facts, to folk economic theories" (Antoci et al., 2020, p. 482). In an analytical model they analyze the optimal persuasion effort of a representative influencer who wants to choose one of two competing narratives in order to maximize his welfare which depends on his impact on public opinion. The model describes the public opinion dynamics that result from the influencer's behavior and the properties of the diffusion process.

## 2.2 Macroeconomics – narratives and economic fluctuations

Shiller's (2017; 2019; 2020) contribution to the study of economic narratives goes well beyond merely directing attention to narratives and coining the phrase "narrative economics". Compared to the other strands of the literature presented herein, Shiller's idea that viral narratives have a causal impact on the macroeconomy provides a fundamentally different paradigm for the power of narratives than any previous approach in economics. Shiller (2019) focuses on grand economic narratives like the rampant fear of technological unemployment in the 1930s. He works out the conditions under which such narratives can form and what factors determine whether the narrative ends up spreading virally throughout society. The most important aspect of his brand of narrative economics, however, is the notion that narratives ultimately feed back to the macroeconomy by influencing decision making and behavior on a broad scale. In line with his past work on investor enthusiasm and irrational exuberance (Shiller, 2000), examples provided by Shiller are often associated with situations characterized by self-fulfilling expectations such as bank runs, real-estate bubbles and the wage-price spiral. Fertile ground for high contagion rates of narratives is provided by narrative constellations – a combination of similar narratives working in tandem – and high rates of repetition of a narrative.

Collier and Tuckett (2021) describe regional disparities within countries as a narrative problem in need of fixing through collective action. They frame the decline of regions as a coordination failure between different local networks, such as business or public policy, to act in concert as a result of insufficiently shared narratives about the region. To overcome this narrative dysfunction, beliefs need to be reset incrementally by a universally trusted leader who may bridge those networks (Collier and Tuckett, 2021, p. 106). According to this model, narratives serve the role of setting beliefs and informing actions that carry significant and macroeconomic consequences, which corresponds well to Shiller´s foundational ideas about economic narratives. However, the notion of an agenda-setting leader actively promoting a narrative to help it build up to a critical mass is somewhat at odds with Shiller´s ideas on the formation of narratives. While Shiller does describe the connection of a narrative to a charismatic person or a human-interest story as increasing its virality, he does not ascribe a coordinating function to this person. The human-interest component makes the narrative more attractive to remember and easier to tell, but the narrative itself, according to Shiller, forms and spreads bottom-up. Resembling Beckert and Bronk´s (2018) concept of *Imaginaries*, Collier and Tuckett maintain that, over time, narratives can harden and get ingrained in a social identity, eventually graduating to *deep stories* that become harder to change. Collier and Tuckett (2021) also ascribe the term *narrative* to monetary policy makers influencing inflation expectations through careful communication of policies and economic conditions. In the wake of the financial crisis of 2008, some high-profile publications emphasized the importance of popular stories and other previously marginalized aspects of economic life in order to show a path forward for the field of macroeconomics whose blind spots were blatantly exposed by the crisis.

Reinhart and Rogoff (2011) describe how verbal accounts of macroeconomic and societal stability rationalize a feeling of comfort and security that prevails amongst economic professionals and politicians alike, even in the face of uncertainty, instability and looming crises. Even though Reinhart and Rogoff do not refer directly to narratives, the stories they recount fit well into a narrative framework. Like Shiller (2000; 2019), Reinhart and Rogoff (2011) refer to asset price bubbles. But while Shiller examines real estate bubble formation from the perspective of the private household attempting to profit from the (supposed) ongoing rise in prices, Reinhart and Rogoff (2011) focus on how scientific narratives on those developments often foster a sense of acceptance and deny the diagnosis of a serious threat to stability. In both cases, agents use a biased selection of facts and metrics and string together a sound narrative in order to justify their actions and move forward. With regards to financial crises, Reinhard and Rogoff suggest a regulatory scheme that allows politicians and experts to hedge against the downside risk of crises, despite widespread belief that a crisis is improbable. In essence, they accept the enormous convictive power of simple narratives and argue for a political framework that is robust to exuberance instead of trying to curb the exuberant narrative itself. Akerlof and Shiller (2010) reinvigorate the Keynesian notion of animal spirits to explain not only the financial crisis and adverse financial events but economic behavior in general. They highlight the economic importance of concepts such as fear, confidence, a concern for fairness and the spread of popular narratives. These concepts are difficult to formalize and, to varying degrees, contradict the assumption of rationality and have thus been largely neglected in economic theory. Akerlof and Shiller (2010) postulate that economic decisions often hinge on the belief or disbelief in certain stories because stories can influence expectations, inspire confidence or instill fear in economic agents.

In the past decade, text mining techniques have increasingly been utilized to identify topics in an economic policy context, particularly in finance and the central bank communications literature. Not unlike the idea put forward by Akerlof and Shiller (2010), the simple premise of this research is the assumption that verbal or textual information provided by policy makers can influence expectations and, therefore, economic decision making. At the same time, the structure of the information that is extracted from text sources and then used to explain variations in economic variables has not been informed by economic theory (or theory of any other kind) but is largely a result of the available tools.

The benchmark method, Latent Dirichlet Allocation (LDA), is a topic model that identifies term clusters resembling topics from large corpora of documents based exclusively on their co-occurrence in these documents. Such topics can then be assigned to documents probabilistically and their prevalence over time may in turn serve as a basis for predicting economic variables. Authors are now starting to refer to Shiller´s concept of economic narratives in empirical studies. In a paper involving news from the US, Japan and Europe, Larsen and Thorsrud (2018) investigate whether certain growth-related news topics identified with LDA `go viral´ and exhibit cross-country spillovers. Building on Shiller´s epidemiological paradigm, they show that international narrative epidemics tend to be US-centric – ie. they are mostly associated with macroeconomic developments in the US – and generally last for 4 to 5 months. Using a Dynamic Factor Model (DFM), they also find that narratives correlated with economic expansions differ systematically from those associated with contractions and that topical diversity tends to decrease in and around recessions.

Borup et al. (2020) investigate open-ended questionnaires from a daily survey of US investors to quantify the real-time development of narratives related to the economic impact of COVID-19. Using a large VAR-model containing LDA-topics and macro-financial variables, they find that narratives and the macroeconomy exhibit a bi-directional relationship. This finding confirms a similar point made by Shiller (2017) which he views as a fundamental problem for implementing causal research designs in narrative economics. While Borup et al. (2020) make the rather conventional methodological choice of quantifying narratives through LDA, their choice of using survey data is unique since it does not rely on the selection and filtering mechanisms present in the context of news or social media. However, they emphasize that the three aforementioned sources can serve as complements when identifying the diffusion of popular narratives because they influence each other in complex ways. The researchers find different lead-lag relationships between sources that depend on the time horizon of transmission and the content of the narratives themselves. Specifically, *job loss* narratives first appear on questionnaires and then – as labor market statistics are published – spread to the news and social media, reflecting a divergence between subjective fear of job loss and news-based *job loss* narratives. The authors find that these dynamics between sources are atypical and the direction of diffusion tends to be reversed for most other COVID-19 news updates.

## 2.3 Applied econometrics
A small strand of the empirical macroeconomic literature marries a particular concept of economic narratives to the dominant DSGE-framework. It builds on a study conducted by Romer and Romer (2004) who use qualitative, narrative data issued by policy makers to refine standard econometric models of the macroeconomy. To attain better estimates of the effects of monetary policy, they infer *intended* funds rate movements from historical FED-documents, arguing that using these narrative records helps to eliminate the endogeneity inherent in many actual policy changes. The paper has spurred an extensive literature that uses verbal information to classify policy changes (Romer and Romer 2010, Cloyne 2013, Gil et al. 2019, Drautzburg 2020, Mertens and Ravn 2014) and firm behavior (Pitschner, 2020). While the term *narrative* was first applied to this type of research in the 2000s, Romer and Romer started using narrative information contained in FOMC documents to identify monetary shocks much earlier (Romer and Romer, 1989). Li (2017) notes that the narrative method finds government employment compensation to play a dominant role in fiscal policy while increases in government purchases of goods and services tend to dominate according to standard VAR-models.

## 2.4 Sustainability, transition and future thinking
In the sustainability and future thinking literature, a larger tendency for researchers to develop or craft narratives can be observed compared to the other categories. Through this notion of creating narratives, this strand of the literature is conceptually related to the view prevalent in political science that narratives are actively created constructs that also serve a specific political purpose. The underlying concept of narratives adopted in the sustainability literature is a rich one: Rather than using the term *narrative* in an act of mere dissociation from a particular idea or from an interpretation of the facts – as it is common practice in political economy –, research tends to focus on both the consequences of certain narratives for society and the roots and terms of their initial formation.

Blignaut and Aronson (2020) advocate the development of a global restoration narrative to coordinate action against climate change. They proceed from the diagnosis that the lack of sufficient action taken by developed countries against climate change is in large part due to the complexity of societal and ecological systems. This results in ecological restoration turning into a *wicked problem*. Disparate groups of actors with radically different beliefs and frames of reference need to work together to solve this wicked problem which in turn evokes fundamental uncertainty and conflict. To overcome this wickedness and to nurture a restorative culture, a *grand narrative* is needed. Daigneault et al. (2019) develop five narratives with regards to possible developments in the forest sector, each reflecting different degrees of sustainability along the dimensions of climate change adaptation and mitigation. In contrast to the political economy idea of narratives being constructed for a political purpose, the authors construct these narratives in an explicit attempt to inform the climate change research and modeling community. The term "narrative", as used by the authors, roughly approximates the term scenario. Schanes et al. (2019) use narratives as tools for scenario development: Three scenarios for a transformation to a resource-efficient society are developed that are characterized by different principal actors and governance models. Terzi (2020) reviews literature from a diverse set of disciplines to shed light on how an impactful behavioral narrative for decarbonization could be crafted. This attempt also reflects the paradigm that large-scale buy-in by the public and fundamental lifestyle changes need to occur to achieve a green transition.

Miedzinski (2018) presents a tool for the analysis and creation of narratives emerging from the political process. He tries to identify and reconstruct storylines voiced by political actors that are related to future scenarios and outcomes. His analytical framework POLiFRAME imposes four layers on such narratives: First-order problems (the current issues that need to be addressed), systemic deficiencies (the roots underlying those issues), scenarios of change (solutions to overcome the issues) and future vision ("What does the world look like absent of these issues?"). These narratives span several temporal dimensions from the interpretation of historical facts to a prospective policy vision. The combination of these dimensions and possible incoherencies between them allows for an analysis of the actor´s deeply held worldviews and assumptions. The goal of this framework is to allow for a critical reflection on policy frames. The POLiFRAME methodology of separating narratives into several layers of understanding has strong conceptual ties to causal layered analysis (CLA) (Inayatullah, 1998), a qualitative method for scenario planning that attempts to enhance and structure our understanding of the forces and ideas that shape different futures. However, studies that instead place their focus on analyzing narratives still form a majority.

Raven and Elahi (2015) develop an analytical model of narrative structure in futures research using insights derived from narrative theory, motivated by the premise that storytelling is often an integral ingredient of scenario development. They demonstrate how, in futures projects, the literary terms *world*, *story* and *narrative* can be considered conceptually equivalent to *data*, *analytical approach* and *final outputs*, respectively. According to their arguments, a narrative developed by scenario developers can be susceptible to be tarnished – either accidentally or purposefully – by manipulative rhetorical framing, resulting in the undermining and blunting of its impact. Borrowing from a range of narrative theories including Plato´ poetic dichotomy, Raven and Elahi (2015) maintain that avoiding these issues requires matching the narrative strategy and mode of a research paper to its purpose. Wittmayer et al. (2019) exploit the linkages between future studies and narratives, describing the *narratives of* *change* (NoC) employed by social innovation initiatives and highlighting the role of such narratives in forming social identities and collectively shared worldviews. Coulter et al. (2019) conduct a qualitative study by interviewing climate change professionals, finding that a lack of shared future-oriented change narratives precludes proactive climate change adaptation measures from getting realized.

Researchers in this literature are conscious of the fact that narrative as a mode of cognition is intimately linked to transformative thinking. Liveley et al. (2021) emphasize the importance of narratives for future thinking, arguing that, since narrative theory aids in recognizing cultural perspectives and also the limits of human imagination, working with narrative tools is vital for enhancing future literacy. Their argument highlights a fundamental connection between sustainability and narratives: The former necessitates developing insights and ideas about an uncertain future while the latter is able to evoke, formulate and structure those ideas. However, the connection between narratives and sustainability extends beyond this relation: Since sustainability studies look for ways to enable a transformation of society by certain criteria, they tend to transcend positive science to also involve normative judgements. Narratives can express these normative evaluations in a clear and concise manner.

## 2.5 Microeconomics and decision theory

Modern decision theory has overcome the monoculture of rational choice theory long ago: Since the inception of the subfield in the 1950s (Simon, 1955), behavioral economists have been working out a laundry list of biases and heuristics, explaining behavioral anomalies challenging the hegemony of the mathematically convenient but unrealistic rational choice theory (Kahneman and Tversky 1979, Thaler and Shefrin, 1981).

Callahan and Elliott (1996) mark an early – if not much-noticed – call to integrate narrative into behavioral economics and thus put the content of human deliberation into focus in addition to its processes and outcomes. Maligning state-of-the-art experimental designs as too restrictive and considering Etzioni´s (1988) argument that norms and affects are central to decision making, the scholar´s call for the use of free-form narratives to further the study of real-world behavior. Much later and firmly in the tradition of the highly influential descriptive approaches to decision making, Tuckett and Nikolic (2017) develop the conviction narrative theory (CNT). According to the CNT, economic actors constantly navigate an environment of radical uncertainty by building plausible narratives about the consequences of their actions. Through interactions with their social environments, conviction narratives are formed that allow actors to weigh their options, make a choice and stick to their decision for the long run. Bénabou et al. (2018) utilize a conception of narratives that builds on the epidemiological analogy proposed by Shiller (2017) and apply it to the choice to behave either selfishly or altruistically in a social context. The focus of their analysis concerns the spread of two contradictory narratives through a network – one narrative promoting responsible and prosocial, the other selfish action that disregards negative externalities. Depending on the interaction structure within the network, prosocial or antisocial norms can emerge as the de facto moral standard and form a behavioral equilibrium.

Narratives are also analyzed in experimental settings. Yang and Hobbs (2020) compare the efficacy of using narrative information instead of logical and scientific reasoning to elicit a positive attitude from consumers regarding the use of gene editing in food technology. They find that telling a story using a personal and personable style significantly improves acceptance of novel biotechnology compared to conveying scientific information in a matter-of-fact way. In the tradition of the framing-effect first described by Kahneman and Tversky (1979), narrative is used as an alternative way of conveying, or framing, information. Harrs et al. (2021) also analyze the effect of narratives in a behavioral experiment. Their conception of narrative goes well beyond how information is presented and instead describes differences in judgment. Optimistic and pessimistic assessments of the COVID-19 pandemic are presented to subjects and are found to have a significant impact on the subjects´ expectations regarding the economic consequences of the pandemic.

## 2.6 Definitions and paradigms used in the literature

The aforementioned literatures differ substantially in how they use the term *narrative*. A crucial paradigmatic difference arises with regards to the role narratives are assumed to play in the economy. Most of the literature under study does not assign an active role to narratives for economic processes. Three basic functions of narratives can be recognized: (1) Narrative as "the bottom line" or an interpretative summary of the facts and mechanisms that explain an issue, (2) narrative as a medium for policy analysis, (3) narrative as an *active* driver of the economy.

The paradigm that is most widespread in the literature considers a narrative to provide an *interpretative summary of the facts* concerning a particular issue – often controversial and political – that can be scientifically described, argued and – in some cases – actively created by the researchers. This viewpoint is particularly virulent in the political economy and development economics literature. Here, the label "narrative" is usually attached to a stance considered to be true by most. Tagging such a viewpoint as a "narrative" is often done in order to attack its hegemonic status, enable criticism and allow for the proposition of an alternative, often contrary narrative.

The second basic paradigm views narratives as a *medium for policy analysis*, as a particular kind of information that can help to broaden understanding about political decision makers and their motivations for policy changes. This gain in understanding can then be used to categorize policy decisions, refine econometric models and improve the identification of policy-relevant parameters. This conception of narrative is manifest in the applied econometrics literature presented in 2.3 This concept of narrative is flexible and simplistic. It essentially reduces a narrative to be a set of policy-related information that is deemed relevant for a study. The narrative is important only insofar as it provides insights into policy making. It does not function as an independent driver of economic processes. If anything, the passivity of such a narrative is a necessary condition since an independent economic effect of the narrative would bias the parameters of interest in these studies.

The conception of narratives as *active drivers of the economy* is virulent in the macroeconomic literature described in section 2.2. This narrative concept is both very rich and also views narratives as central to economic outcomes. This paradigm is also adopted by Tuckett and Nikolic (2017) and Collier and Tuckett (2021) as a fundamental assumption of CNT in the context of microeconomics and decision theory in 2.5. Tuckett´s and Shiller´s narratives can be considered as two sides of the same coin, the former laying out the individual cognitive conditions for the development of narratives and the latter describing their social diffusion and macroeconomic consequences. More than one of these functions can be expressed or implied in any piece of research. However, this third, *active* function of narrative is specific to narrative economics as it implies a paradigm shift. Narratives are not reduced to transporting information *about* the economy, but they are endogenous drivers of activity *in* the economy.

# 3 Definition of the concept "collective economic narrative"

Our literature review shows that the term *narrative* is used with different meanings in economics and that the provided definitions are often rather vague. In many cases, economist authors do not even define the term, but rather give a loose description. We believe that conceptual clarity is important for scientific progress and hence propose a definition of *economic narratives* that captures important aspects mentioned in different strands of the literature. We distinguish *collective* narratives from *personal* or *private* ones, which people create for themselves and by themselves, e.g. to make sense of their own life and to create their personal identity. "Collective" means that the narrative has relevance and functions in a social context. Our definition is meant to be relevant and fruitful for further research in *economics*. We do not want to make any claims about the usefulness of our definition in general.

We derive our definition from a complexity perspective on economics, which emphasizes off-equilibrium dynamics, novelty and adaptation and regard the economy as a complex adaptive system (Roos, 2017). According to Arthur (2015, p. 24) complexity economics "sees the economy not as a system in equilibrium but as one in motion, perpetually `computing' itself – perpetually constructing itself anew. Where equilibrium economics emphasizes order, determinacy, deduction, and stasis, this new framework emphasizes contingency, indeterminacy, sense-making, and openness to change". Note that Arthur mentions *sense-making* as an important element of complexity economics. Agents in a complex adaptive system can have only very limited knowledge of the whole system in which they act. Their knowledge is always preliminary, subject to change and constructed in social processes (Richardson, 2005). Narratives play a crucial role in these social processes of knowledge generation and sense-making. They enable agents to act purposefully in an uncertain and ever-changing environment (Tuckett and Nikolic, 2017).

The complexity perspective is well compatible with literature from institutional economics[i], e.g. Denzau and North (1994), Searle (1996, 2005), Hodgson (2006) or Dolfsma et al. (2011), which is also highly relevant for economic narratives. Searle (1996, 2005) proposes the notion of language as the fundamental institution, which is necessary for the development of all other institutions. According to Searle most economic reality consists of institutional facts, which only exist due to the collective acceptance of something X having a certain status Y. A status carries functions that give its bearer X the power to do something. For example, a banknote X has only the status of money Y, because everybody assigns this status to it. If a banknote has the status of money, it gives its owner the power to buy something with it. Collective acceptance must be linguistic or symbolic, because the status function is not a physical property of X, but must be represented in the minds of individuals and communicated between them. In this sense, Searle considers language as the fundamental social institution. Narratives are linguistic means to assign the status functions. Walsh and Stepney (2018) establish a link between complexity theory and narrative theory.

The complexity-cum-institutional perspective appears promising to us, because it allows us to draw upon a host of previous research in different fields dealing with narratives and similar concepts. We relate our definition to this literature below[ii].

We propose the following definition:

*A collective economic narrative is a sense-making story about some economically relevant topic that is shared by members of a group, emerges and proliferates in social interaction, and suggests actions.*

Next, we explain the elements of the definition, argue why we consider them important, and relate them to a large body of literature from different disciplines.

### 3.1 Story

It is uncontroversial that narratives are stories. Many authors even use the terms *story* and *narrative* interchangeably without ever really defining what a story is. We suggest using *story* to mean an articulation of a temporal sequence of events. The literary theoretician Gerald Prince (1973) defines an event as any part of a story that can be expressed by a single sentence. This implies that a story has a certain structure and that a simple collection of terms or categories is not a story (and hence not a narrative).[iii]

The English novelist Edward Morgan Forster (1927) gave an example of a (minimal) story: "The king died and then the queen died" (Forster, 1927, p. 86). The merit of a story is that the reader or listener wants to know what happens next, i.e. a story appeals to the curiosity of the audience, which he considers "one of the lowest of the human faculties" (Forster, 1927, p. 86). Forster contrasts a story with a *plot*: "`The king died, and then the queen died of grief' is a plot. The time-sequence is preserved, but the sense of causality overshadows it" (Forster, 1927, p. 86). A plot provides an explanation why events unfolded in a temporal sequence. An example of a minimal economic story is: "Inflation rose and then the central bank raised the interest rate". A second story would be: "The central bank raised the interest rate and inflation slowed down." Note that the reader might infer a causality between the mentioned events, but the stories themselves do not mention a causal relationship.

For Prince (1973), however, two conjoined events are not enough for a minimal story. He requires a minimal story to consist of three events. The first and the third event describe states and the third event is the inverse of the first, e.g. "Korea was poor" and "Korea is rich". The second event describes an action that causes the third event, for example "Korea invested in education". According to Prince's definition, the two central bank stories mentioned before are just one: "Inflation was high. Then the central bank raised the interest rate. As a result, inflation is lower now."

Note that neither Forster nor Prince explicitly mention actors as part of a story, although many of their examples contain actors. Actors can enter stories indirectly as subjects in simple clauses. In the English language, a complete simple sentence includes a single clause, which comprises a subject and a predicate.

Narratology or narrative theory is the discipline specialized on narratives, whose origins reach back to Aristotle's Poetics. It is impossible to give an overview in this paper, but some remarks on narratology seem relevant. In general, narratology examines how narratives and narrative structure affect the human perception and understanding of the world. There seems to be an intricate relation between narrative, language and the mind. Ferretti (2022) argues that the possession of a "narrative brain" distinguishes humans from other animals and gives them a huge evolutionary advantage. Having a brain that allows humans to tell stories makes cooperation in large groups possible. While classical narratology aimed at a taxonomy of the fundamental elements of narratives and their rules of combination, cognitive narratology draws on neuroscience and cognitive science in order to understand the interaction between narrative and the human brain. Armstrong (2020) argues that one approach in cognitive narratology builds on the structuralist paradigm linking narrative structure to universal structures of mind such as *frames* or *scripts*. In this view, narrative structure follows from the underlying mental structure. A second approach rejects the notion of universal rule-governed structures of mind, language and narrative and emphasizes that embodied minds, stories and the world interact. In this view, readers have bodily experiences to which the mind responds. Breithaupt (2022) argues that the brain of a human baby must learn to structure and order the vast stream of sensory information. The formation of closed units is a fundamental principle of creating order. Stories organize events in time. Similar to Prince (1973) Breithaupt argues that narratives are closed units or episodes, if they consist of three parts: a beginning, an end and a middle part. The end is especially important, because it closes the unit and signals to the brain that it can direct its energy and awareness to other things.

In a general sense, the aforementioned structural aspects that characterize the story make it easier to comprehend and remember than expository texts such as essays (Mar et al., 2021). It is also insightful to consider the research on the effects of fiction, i.e. imaginary stories, on the brain. When we comprehend fictional narratives, we rely on similar cognitive processes we use when we make sense of real-world events (Gerrig and Mumper, 2017). According to Oatley (2016), "fiction is the simulation of selves in interaction. People who read it improve their understanding of others" (p. 618). The concept of parasocial relationships describes the immersive process of forming imaginary relationships with media personae, among them fictional characters. It is one of the major pathways in which fiction exerts influence on our perception and attitudes toward ourselves and other people in our lives (Brown, 2015). While the specific neural mechanisms at play have only recently begun to be studied (Broom et al., 2021) and some studies also find certain differences in the processing of factual vs. fictional information (Altmann et al., 2012) , the mental processes of engagement with fiction appear to be similar to those of understanding other people. This may be because both processes require changes of perspectives. If fiction can then be thought of as a kind of simulation of the social world, consuming fiction might improve empathy, theory-of-mind and social skills (see also Bruner 1986, Mar and Oatley 2008, Mar 2018, Jacobs and Willems 2018).

### 3.2 Sense-making

A narrative is a special kind of story. The sense-making characteristic of narrative implies that the story has a deeper meaning for the speaker and the listener, one that transcends the level of the events discussed in it. It is told with the intention to understand the world and to interpret some data, event or action. In this paragraph, the focus is put mainly on sense-making in the sense of interpreting events and their mutual connections to form a sense-making story. However, sense-making is a multi-level process that extends to much finer granularity than the connection of given events. Event Segmentation Theory deals with the more fundamental question of how a sequence of states perceived by the senses can be divided into events (Zacks et al., 2007). Generally, sense-making has a cognitive, an emotional[iv] and a normative component.

The cognitive component refers to the assignment of causal relationships between events. Something can make sense for us if we can provide a reasonable explanation why it occurred, i.e. how it was caused. Recognized patterns and the knowledge of causal relationships also make it possible to predict what is likely to happen after a certain event has been observed. The ability to predict future events is important for making decisions. In a social context, sense-making can also mean that a justification for behavior is given. A sense-making story can explain why an agent has behaved in a certain way in the sense of providing a justification for it. Narratives as sense-making stories are plots in the language of Forster (1927). Therefore, a "because"-sentence could be a minimal version of a sense-making story, e.g. "Inflation slowed down because the central bank had raised the interest rate". Sense-making stories related to the future could be if-then statements like "*If* the central bank raises the interest rate, *then* inflation will slow down".

Shiller (2017) has speculated that emotional content of economic narratives may contribute to its virality. The presence of emotion indeed exerts a substantial influence on how information is perceived and whether or not it is retained in memory (Tyng et al., 2017). As a result, emotionally charged information has a competitive advantage in neurological processes (Pessoa, 2013). Especially in a setting where multiple stimuli compete for cognitive resources, emotional information is prioritized and has a greater chance of generating attention (Vuilleumier, 2005) and motivating behavior (Tyng et al., 2017). Because the modern media environments in which economic narratives are conveyed are noisy and attention-driven, this kind of privileged access to our neural capacities is highly relevant. In the end, the emotional component of sense-making may show up as intuition or a gut feeling. We may have a rational explanation for something, but it may feel wrong. We can also have an intuition about something without being able to explain it rationally. Either way, emotions alter how plausible the causal conclusion of a narrative feels to us (Tuckett and Nicolic, 2017).

The meaning of a story can also result from an evaluation of whether the topic of the story is right or wrong in a normative sense. Especially when we hear a story about something that we consider normatively wrong, we might be motivated to do something about it. The cognitive, emotional and normative elements of sense-making are interwoven and hard to separate. Sense-making requires the story to connect to the *belief systems* of the people involved. We use the term belief system in a broad sense here that includes *mental models* as well as normative, evaluative, affective and motivational elements (Abelson, 1979). Mental models are cognitive representations of the external world, which people use to interact with the world and to make decisions (Jones et al., 2011). They describe the entities people perceive to exist in the external world and the relationships between them. People use them to understand and explain what they observe in the world and to anticipate the future. However, people do not only form representations about how the world works, but also hold beliefs about whether this is good or bad and how the world ought to be. Belief systems include representations of alternative worlds in addition to the mental models of the existing world, which can motivate them to take action in order to change the existing world (Abelson, 1979). We call the evaluative and prescriptive part of the belief system the *value system*. According to Schwartz (2007), values are beliefs linked to affect which motivate action and serve as standards or criteria. Importantly, values form a system of *value priorities*, i.e. they are ordered by importance relative to one another. Individuals or social groups differ in their value priorities. In our conception, the value system also contains *evaluation rules* by which values are related to the mental models and *evaluations*, which are stored outcomes of evaluative processes.

Personal narratives can be seen as parts of the belief system of a single person. Social narratives, however, involve the belief systems of at least a teller and a listener of the story. The teller may want to share his/her view of the world with the listener. Whether this can succeed depends on the degree of overlap between the belief systems of the two agents. If the belief systems are too different, a narrative may make perfect sense to the teller, but not at all to the listener. A good example for incongruent narratives based on different belief systems is the claim of the Turkish president Recep Tayyib Erdoğan that inflation will slow down, if the Turkish central bank lowers the interest rate, which does not make any sense from the perspectives of most economists[v]. The fact that most economists believe in the opposite does not mean that they must be right, since it is very hard to definitively prove causality in economics. But the point here is that Erdoğan's claim is incompatible with the mental models of mainstream economists, which are derived from mainstream economic theory.

A narrative may transport *subtext* in addition to what is said explicitly and directly. The subtext appeals to the underlying belief system. The use of subtext simplifies the communication of difficult contents among agents who share the same belief system, because only the most important elements are articulated whereas the connections between them, their meaning and the relation to other topics are just implied and must be completed by the listener. At the same time subtext generates ambiguity, because it must be read between the lines. Listeners with a belief system that differs from the one of the teller, either have a different interpretation of the subtext or do not even notice that subtext is present. An example of subtext is when economists speak of "financial shocks" as the cause of financial crises. For the non-economist listener, a "shock" might be interpreted as a "sudden disturbance", but in the DSGE models of macroeconomic theory a shock is an unpredictable stochastic disturbance. Whenever economists who are used to thinking in terms of DSGE models talk about financial crises being caused by financial shocks, they always imply that the crisis was at least partially unpredictable and hence unavoidable.

Note that our definition of narratives as sense-making stories differs from the concept of Eliaz and Spiegler (2020). For them "narratives can be regarded as causal models that map actions to consequence". We distinguish between the causal (mental) model and the narrative and argue that the narrative is a partial articulation of a more complex underlying causal model. If someone says "The central bank raised the interest rate and then the economy fell into a recession due to the bank lending channel", we would consider this a sense-making narrative, but not a complete causal model. The "bank lending channel" is a term for the model that links the increase in the interest rate and the recession, but it is not the model itself. The narrative makes only sense to someone who knows what "bank lending channel" means, i.e. who has an idea of the agents involved, the relationships between them and the assumptions about their behavior.

## 3.3 Shared by members of a group

In order to be a collective narrative, the story must make sense to members of a group and be told within as well as by the group. In sociology and social psychology, a social group is defined as consisting of more than two interacting people who share some characteristics and have a sense of unity, i.e. consider themselves a group (e.g. Reicher 1982). The self-reflective feature that the members of a group think of themselves as belonging to a group is key to the social identity approach (see XXX). Dolfsma et al. (2011) emphasize that a group or community identity requires a common language. Collective narratives can be part of such a common language. A group can differentiate itself from other groups by telling its narratives to outsiders who might have different (competing) narratives: "A social narrative can bind people together since … it is not a single narration event, but a series of narration events through which a story or its versions are retold and reheard, time and time again, by individuals, organizations, or institutions" (Shenhav, 2015, p. 58). By retelling the narrative within the group, a group identity is formed and the belief systems of the members are aligned.

A story can only make sense to a group, if the members of the group share a belief system, at least partially. If either the mental models or the value systems of the group members differ too much, the group cannot share a sensemaking narrative. In fact, the group may be defined by the shared belief system. The shared belief system is not only a prerequisite for the existence of a shared sensemaking narrative, it can also be formed by narratives. By telling each other individual narratives, the members of a group can find out where their belief systems are congruent and where they differ. Forming a group identity can mean that the members of the group adjust their belief systems such that they match better to those of their peers.

The notion that shared belief systems are a prerequisite for collective narratives is very similar to the concept of "shared mental models" as a basis for communication and cultural learning by Denzau and North (1994). Ideologies and institutions can be regarded as classes of shared mental models. Denzau and North explain that shared mental models provide concepts and a common language that facilitate communication. They also emphasize that the mental models of group members tend to converge over time due to communication. Without communication, individual learning would lead to divergence of the mental models of different people, because each individual adjusts their own mental model to private experiences. Communication and the creation of ideologies and institutions are hence co-evolutionary processes. The discussion in Dolfsma et al. (2011) on institutional durability and vulnerability is in a similar vein. On the one hand, institutions are "durable systems of established and embedded social rules that structure social interactions" (Hodgson 2006 p. 424), i.e. they have a certain stability over time. According to Searle (1996, 2005), language as a fundamental institution plays a critical role for institutional stability, because the (re)production of institutions occurs through recognition within a community which is only possible through language. On the other hand, institutions are vulnerable in the sense that they can change, when they are irritated by information from outside. Furthermore, following Luhmann (1984) Dolfsma et al. (2011) argue that even with a common language (or a shared mental model) communication can be imperfect, because different individuals can interpret the same message in different ways.

Other related concepts are "collective representations" (Durkheim 1912) and "social representations" (Moscovici 1961) from sociology and social psychology respectively. Collective representations are forms of knowledge shared by members of a society that help to order and make sense of the world. They also delineate cultures from one another. Depending on their extremity, irreconcilable differences between those frameworks may render cross-cultural communication impossible and consolidate long-term conflicts between societies. Collective representations are mental constructs but they also have physical representations that can be visualized in studies of the brain (Turner and Whitehead 2008). Similarly, social representations are a "system of values, ideas and practices with a twofold function: first, to establish an order which will enable individuals to orientate themselves in their material and social world and to master it; and secondly to enable communication to take place among members of a community by providing themselves with a code for social exchange and a code for naming and classifying unambiguously the various aspects of their world and their individual group history" (Moscovici 1973, p. xiii). Importantly, social representations are both the process and the result of constructing social reality.

Similar to the interaction between narratives and the brain by which the brain adjusts to narratives and narratives are influenced by the structure of the brain, there is multiple causal interaction between narratives, groups and their shared mental models or social representations. Narratives are told by groups, but they also affect the group identity and composition and how the group views the world. This kind of circular causality is a key characteristic of complex systems, such that there is a natural connection between narratology and complexity science (Walsh and Stepney 2018).

### 3.4 Emerges and proliferates in social interaction

Collective narratives are not *created* by any single agent, but are the outcome of repeated interaction between members of a group or of different groups. Collective sensemaking is an interactive process between members of a group, who share similar belief systems. Nobody who tells a story to others can know exactly which parts of the story make sense to them, because it is impossible to observe the complete belief system of others. Furthermore, the originator of a story does not know how and to whom the story is retold. A story proliferates in a group, if the initial listeners like it and tell their versions of it to others. The original story that somebody tells might resonate with other people or not, depending on whether it connects with their mental models and values. The resonating parts of the story are retold and the less convincing parts are dropped or modified. If different versions of a story circulate in a group, the group might try to integrate them into a consensus version that contains the core of the different variants. In the end, a group narrative is left that nobody thought of in this way and that nobody could predict, hence it emerged.

Narratives do not only emerge as a consequence of within-group interaction, but also result from interaction between groups. The narrative of a group might be challenged by other groups that maintain different belief systems. If different groups compete in some arena, they may have an incentive to differentiate their narratives as much as possible instead of aligning them. The evolution of a group narrative depends in complex ways on the participants and the rules and practices of the inter-group discourse.

Akerlof and Shiller (2010) and Shiller (2017; 2019) compare the spreading of stories with virus epidemics. We argue that this view is too simple and that there is a fundamental difference between a virus epidemic and the propagation of narratives. The transmission of a virus happens unconsciously and passively. Normally, people do not pass on a virus on purpose. In contrast, telling narratives is an active process that can have an intention. While there are stories which are told without a purpose, many agents tell narratives with the intention of influencing others and persuading them to perform certain actions. This is the case for political narratives, which aim at political support by voters, or narratives in a business context, which can be geared toward employees, customers or investors. Often, it is not by chance that narratives suggest action. On the other hand, recipients of narratives may be exposed to them passively similar to a virus, but they might also seek them actively and choose how they react. While getting ill from a virus is not a choice, people at least in principle can choose whether they perform the suggested action or not. In a similar vein Centola (2018) argues that the spread of behavior is different from the diffusion of a virus in a population. He distinguishes simple informational and viral contagions from complex contagions of behaviors such as cooperation, marriage practices, health behaviors or investment decisions. The relevant characteristic of complex contagions is that the transmission requires contact with multiple adopters, while a single contact with an infected person is sufficient to transmit a virus. The change of behavior or the spread of complex information require social confirmation, because they entail some kind of cost, risk or complementarity. In order to overcome these barriers, reinforcement from the social network is necessary for a person to adopt the new behavior. This mechanism is relevant for collective narratives in our conceptualization, too. Shiller might be right that an entertaining story told for fun at a party might spread like a virus. However, narratives as sense-making stories based on shared mental models that influence peoples' behavior are complex objects, whose transmission is likely to require multiple interactions with different people.

### 3.5 Suggests actions

In economics, narratives are interesting because they suggest actions to economic agents. Social scientists analyze narratives not as an object that is interesting by itself, but because they have a function in groups or social systems. One of the main functions of narratives is to enable groups to act despite fundamental uncertainty about the future (Beckert and Bronk, 2018). In particular, they coordinate group action. If actions are interdependent, coordinated behavior is often more beneficial than isolated action. Coordination is achieved, if all members of the group have similar expectations about the outcomes of actions and if they evaluate the outcomes in the same way. This is not achieved by the told story alone, but by the reference to the shared belief system of the group members. The belief system uses the input from the narrative in order to simulate consequences of actions and to evaluate them.

Suggesting what to do in an uncertain world is another interpretation of sensemaking, in addition to the explanation of observed phenomena. Especially in uncertain situations, people often have a desire to act, even though the knowledge basis for rational decision-making is rather small. Patt and Zeckhauser (2000) termed this impulse to act *action bias.* Sometimes, rational behavior can be to do nothing and to wait until new information has resolved part of the uncertainty, but people's self-perception as an actor or decision-maker who has control over uncertainty can urge them to do something. This idea has some similarity to Keynes' concept of animal spirits – defined as "spontaneous urge to action rather than inaction" (Keynes, 1936/2018, p. 141). Akerlof and Shiller (2010) see *confidence* as part of Keynes' animal spirits and argue that stories can create or destroy confidence of consumers and investors. As a consequence, "we must understand that saving depends upon the stories we tell about our lives and our future" (Akerlof and Shiller, 2010, p. 119). As one of many examples they mention the book "How a second home can be your best investment" (Kelly and Tuccillo, 2004) which is filled with personal stories of people who bought homes and got happy. Akerlof and Shiller argue that such stories served as models for many people's behavior and hence contributed to the housing bubble in the U.S. before the financial crisis 2008. Similarly, Reinhart and Rogoff (2009) argue that before every financial crisis, financial analysts tell stories why "this time is different", i.e. conventional standard of asset valuation no longer apply, and why investors should carry on buying assets.

## 3.6 Collective economic narratives: the bottom line

According to Robert Shiller (2017), the key question in narrative economics is why some narratives are more engaging and effective in ''going viral'' than others. The definition discussed above tries to work out features of narratives that make them capable of influencing decision making on such a large scale that they end up changing the course of the macroeconomy. This procedure, however, should not obscure the fact that, in reality, the five parts of our definition will not at all operate in complete isolation from one another. Modern neuroscience has gradually moved away from trying to localize specialized brain regions that are only responsible for – say – emotion, perception or social cognition. While those categories are still often talked about in everyday conversation as if they were completely separate entities, they are probably processed in an integrated way through domain-general networks in the brain (Barrett and Satpute, 2013).

Therefore, it is easy to find links and points of overlap between elements of our definition: For example, the subtext of a narrative connects sense-making (3.2) to its spread within a social group (3.4) and – ultimately – a shared sense of reality in that group (3.3). Since only group members will agree with or even understand the subtext of a narrative that is told within it, a group identity is forged and maintained through the common sense of truth that the subtext contains. Some things are so fundamental to us that they literally "go without saying". Stories (3.1) and prosocial acts (3.3 and 3.5) are also inextricably linked: The structure of a story – characters who engage in a sequence of events – prompts our brain to perform a social simulation (Oatley, 2016): "What would I do if I ended up in this situation?". Even the medium of this simulation is somewhat understood: The neurotransmitter oxytocin has been found to encourage socially cooperative behavior. It is also emitted when people are exposed to narratives. More so, the degree of prosocial economic behavior that people are willing to engage in increases with rising oxytocin levels in the blood (Zak, 2015). This means that there is a line to be drawn between the story structure (3.1), emotional sense-making (3.2), social attitudes (3.3) and, finally, behavior (3.5). The dynamics between the stories told within groups and that group´s shared mental models are driven by a two-way causal connection. Collective narratives emerge and take shape as a result of interactions within groups, but those narratives in turn affect group composition and how the group views the world. This kind of circular causality is a key characteristic of complex systems, providing a natural connection between narratology and complexity science (Walsh and Stepney 2018).

According to Ferretti (2022), the "narrative brain" – a brain capable of compressing information into narratives – is the prerequisite for the narrative to emerge as a communicative tool. Narratives, according to this view, appeal to our brain because they fit neatly into its preferred way of processing information. But irrespective of the narrative *brain* or narrative *communication* coming first, narratives "provid[e] an *economical* cognitive instrument for understanding everyday life" (Robinson and Hawpe 1986, p. 113, emphasis added) by compressing the overwhelming complexity of information that we can perceive about a matter into a neat, consistent package that we can act on.

Differentiating collective economic narratives from other kinds of information is a major challenge for economists. Figure 3 provides a taxonomy of textual forms that may serve as a "checklist" in the classification of texts and also provides several precursors to *collective economic narratives*, should one or more of the criteria not be satisfied.

**Figure 3: Is a text a collective economic narrative?**

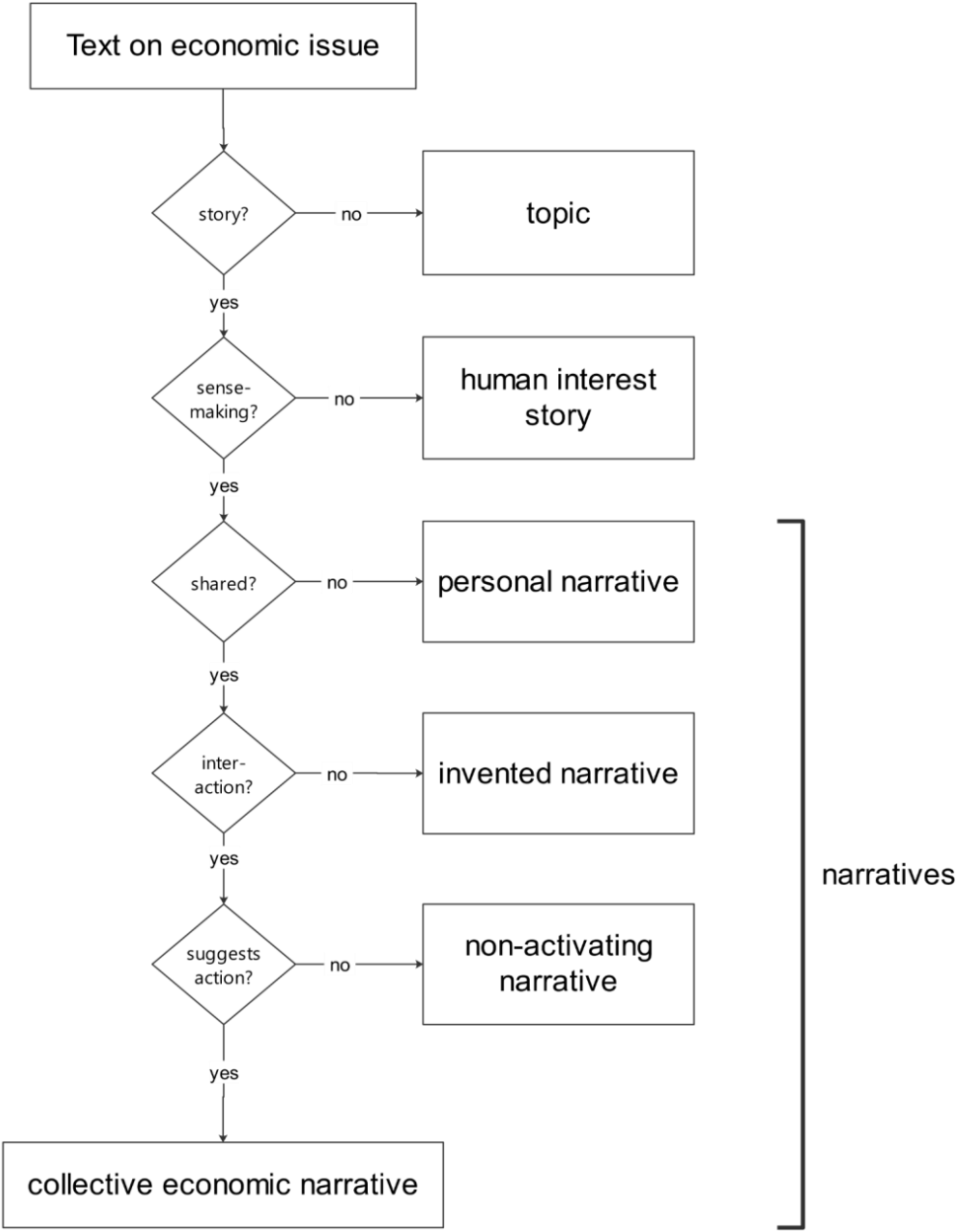

## 4 Examples from the literature

In this section we discuss two examples of how the term *narrative* is used in the current macroeconomic literature. Our objective is to show that the current use of the term in the economic literature is very imprecise. In both examples, "narrative" just means "topic". Our first example is what Robert Shiller calls the Great Depression narrative in the context of the financial crisis 2007 - 2009. We argue that Shiller does not really identify a narrative, but a discourse in which several narratives were used. We suggest formulations of three narratives that are in line with our concept presented in the previous section. The second example shows an attempt to identify narratives with statistical topic modeling. While the macroeconomic analysis of the detected topics is interesting, it does not exploit any features of narratives.

## 4.1 Shiller's Great Depression narrative

Shiller (2017) provides examples for his concept of economic narratives that have a macroeconomic impact. One of them is the *Great Depression narrative* between 2007 and 2009 in the U.S.: "The 2007–2009 world financial crisis has been called the Great Recession as a reference to the Great Depression of the 1930s. Certainly, the narrative of the Great Depression was suddenly thrust into the national attention as never before, not since the 1930s" (Shiller, 2017, p. 994). As an illustration for the existence of this narrative, he presents the frequency of appearance of the phrase "Great Depression" in news, newspapers and books, which skyrocketed in 2007.

Referring to Google Trends search counts, he argues that people were not really interested in the details of the events in the 1930s, because terms related details of history were not searched a lot. "It was more just a quick and easy way to communicate narrative: we have passed, by 2007, a euphoric speculative immoral period like the Roaring Twenties, the stock market and banks are collapsing in 2008 as around 1929, and now the economy might really collapse again like that; we might even be unemployed and on the street crowding around failed banks, yes really! End of basic narrative" (Shiller, 2017, p. 994).

We argue that Shiller does not describe a Great Depression narrative, but a *public discourse* about the Great Depression. More precisely, the discourse is not about the Great Depression itself, but about how similar the situation in 2007 – 2009 was to the situation in 1929 – 1934 and what the past meant for the present. The crisis of 2007 – 2009 was of course an economically relevant topic and it seems safe to claim that people wanted to make sense of the events that occurred. The stylized story that Shiller presents mentions apparent similarities between the two time periods, suggesting that people wanted to learn from history about their current situation, which is a case of sense-making. The story does not suggest certain actions, which is part of our definition of narratives. Discourse means that there are several competing narratives about the same topic, each of them shared by different groups and proposing different interpretations and suggesting different actions. We cannot present a systematic analysis in this paper, but we claim that there might have been at least three different Great Depression narratives at the time[vi]:

1. *Fiscal stimulus is needed to prevent another Great Depression.*
2. *Monetary policy has learned the lessons from the Great Depression.*
3. *Elites forgot the lessons from the Great Depression.*

The first narrative is mentioned by Shiller himself: "In 2007–2009 presidents and prime ministers invoked parallels to the Great Depression to justify their requests to apply stimulus" (Shiller, 2017, p. 996) and "During the 2008–2009 financial crisis politicians around the world warned of the risk of an imminent depression in a bid to win acceptance of aggressive stimulus policies" (Shiller, 2020, p. 796). A three-event narrative, following the scheme of Prince (1973), might sound as follows: "The depression of the 1930s was prolonged by fiscal policy that was too passive in the beginning. Keynesian theory, which was developed as a consequence of the depression, demonstrates the benefit of expansionary fiscal policy in a liquidity trap. A long and deep depression following the 2007 - 2009 financial crisis can be avoided by strong fiscal intervention." The action that is suggested by this narrative is obvious. The crucial aspect here, however, is not the specific three-event structure, but the sense-making glue between those events.

The second narrative is related to the first one, but told by and about a second group: central bankers. The public noticed quite early during the financial crisis that Ben Bernanke had studied the mistakes of monetary policy in the 1930s as a scholar before he became chair of the Federal Reserve Bank. Hence there was a narrative similar to the story about fiscal policy, but with the twist that it was personalized: "In the 1930s, the central bank contributed to the transformation of the recession into a depression by keeping interest rates too high. Ben Bernanke revealed this mistake by studying the history of the Great Depression. As chairman of the Fed, Bernanke knows what to do in order to avoid the mistakes of the past[vii]". This narrative was shared by policy observers such as Gros and Alcidi (2009, p.2): "The conclusion for monetary policy is clear: the errors of the 1930s will not be repeated (policy) interest rates have been lowered decisively and quantitative easing is being actively considered even by the ECB". It was also told by Bernanke himself. On 8 April 2010 he said in a speech[viii]:

> *"I was an academic economist and economic historian, with a particular interest in the causes of the Great Depression. … I thought that I would speak to you about the parallels--and differences--between that crisis and the more recent one, particularly regarding the responses of policymakers …. For its part, the Federal Open Market Committee, the monetary policymaking arm of the Federal Reserve, sharply and proactively cut its target for short-term interest rates from the fall of 2007 through 2008. … Using emergency authority last employed during the Depression, we created an array of new facilities to provide backstop liquidity to the financial system (and, as a byproduct, coined many new acronyms). Thus, we were able to help restore the flow of credit to American families and businesses by shoring up important financial markets, such as those for commercial paper and securities backed by consumer loans."*

The third narrative is in stark contrast to the second and tells about the failure of elites that had not learned their lessons from the Great Depression and hence only made the financial crisis possible. The narrative criticizes both neoclassical mainstream economists and neoliberal policymakers for the regulations of financial markets[ix]. The narrative is: "Following the negotiation if the Bretton Woods system in 1944, the risk of financial crises was low. Over time the experience from the interwar period waned and both economists and policymakers liberalized financial markets. As a consequence, the risk of financial turbulence was high and culminated in the Global Financial Crisis". This narrative is shared by economists who criticize the mainstream of their profession, such as Nouriel Roubini or Paul Krugman, but also by political activists such as Naomi Klein[x]. The suggested action is political reform, either in the form of Obama's new liberalism[xi] or radically anti-capitalist[xii].

Of course, these three narratives are not mutually exclusive and they might not be the only ones told in the Great Depression discourse. One might argue that the first and the second narrative constitute just one narrative about both fiscal and monetary policymakers doing the right thing based on the experiences of the past. But given that there was controversy about the right policy responses and the different interests of the actors, distinguishing two narratives provides a more nuanced view. For instance, there was a debate about the Fed's contribution to the crisis, which could have been at least threefold[xiii]. The first allegation is that the Fed had fueled the housing bubble with cheap money for too long. Second, the Fed did not do enough to monitor and stop the malpractices at Wall Street and, finally, the Fed failed to rescue Lehman. It was in the interest of the Fed and in particular of Ben Bernanke to nurture the narrative that the Fed could have done little to avoid the crisis, but did its best to contain it. The topos of Bernanke as the scholar who had studied the Great Depression lends credibility to this narrative because of the subtext that a respected academic would not act in contrast to his deep knowledge.

Our definition does not only contain the element that a narrative is shared by a group, but also that it evolves over time due to social interaction. This evolution can affect both how the narrative is told and its purpose or function. The initial function of the monetary policy narrative might have been to gain support for new and unconventional monetary policy instruments, which were justified by the extraordinariness of the situation, for which the reference to the Great Depression served as an argument. Later on, the focus of the debate about monetary policy shifted from the Fed's responsibility to fight the crisis to the Fed's accountability for its existence. At the later stage the narrative's function for members of the Fed might have been to exculpate them from the accusation that the Fed's behavior had been a cause of the crisis.

## 4.2 LDA-based business cycle narratives

As discussed in Section 2, it has become quite popular to use text mining techniques such as LDA in order to identify narratives in large text corpora and to analyze their impact on the economy with econometric tools. Larsen and Thorsrud (2018) is a good example for this approach. They refer to Shiller (2017) and "define the term narrative to mean a simple story or easily expressed explanation of events that many people want to bring up on news. The news-based topic modeling approach captures this idea, and allows us to identify what the news stories thematically are about in a parsimonious manner" (Larsen and Thorsrud 2018, p.2).

It is interesting that Larsen and Thorsrud (2018) talk of narratives that they identify with the LDA approach. In their companion paper Larsen and Thorsrud (2019), they use the same method on different data, but speak of topics instead of narratives. In fact, they argue correctly that LDA produces word clusters which are called topics. Two examples of the identified topics are shown in Figure 4.

**Figure 4: Word clouds and topic categorization from Larsen and Thorsrud (2018)**

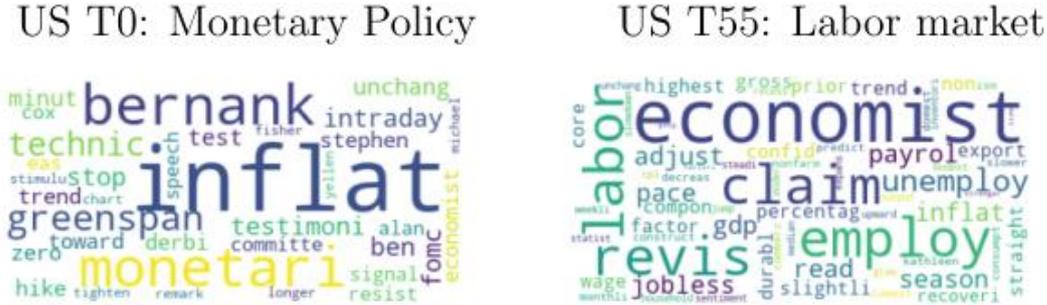

The size of the word in the word cloud generated by the algorithm reflects the probability of this word occurring in the topic. The labels for the topics ("monetary policy" and "labor market") were subjectively chosen by the authors. As visible in the word clouds, the only structure in those topics is the distribution of words. The requirement of a story that a temporal sequence of events is reported, is not met. LDA can hence only discover what the news were about (i.e. the topic), but not the specific contents nor any of the other elements of our definition of narratives. Topic modeling can hence be a first step to the identification of narratives, but not more. Larsen and Thorsrud (2018) make an additional step by translating the topic decompositions into tone adjusted time series, that measure how the news reported about a topic. In order to measure the tone of the reporting, they use a word list with positive and negative words as defined by the Harvard IV-4 Psychological Dictionary.

Using these data, the authors estimate news coincident indexes for the US, Japan, and Europe and show that these news-based indexes track the state of the economy very well. They can also show that some news topics are relatively exogenous and carry information that is helpful to predict the evolution of TFP in the US. However, Larsen and Thorsrud (2018) do not make any claim about the causality between the identified news topics and economic activity.

In a similar fashion, Borup et al. (2020) also claim to identify narratives using LDA, albeit using a different source of data. Although they acknowledge that "a topic is not a narrative in and by itself" (Borup et al., 2020, p. 3), the authors maintain that the essence of a particular narrative can be captured by a set of keywords. We think that these kinds of analyses are valuable and provide insights into the relation between the prevalence of topics and the business cycle. However, they are quite far away from capturing our concept of economics narratives. In fact, the whole empirical analysis could have been done without the term "narrative" and using the terms "discourse" or "topics" instead. The reference to Shiller (2017) and his concept of narratives mainly serves as a motivation and an argument for the dynamics of news stories over time. This is not to say that narratives are not driving the results, but neither Larsen and Thorsrud (2018) nor Borup et al. (2020) really show the relevance of narratives directly. They analyze topics, not narratives.

# 5 Empirical identification of narratives

Our definition of economic narratives is highly demanding for quantitative empirical research. In this section, we discuss some issues researchers face if they want to identify economic narratives with methods of statistical text analysis. Using qualitative text analysis, economic narratives are easier to discover, but such methods are not applicable for an in-sample analysis of large text corpora since they require close reading of texts.

## 5.1 Challenges regarding empirical identification

The minimal requirement for a story is that the text contains a temporal sequence of events. An event means that something happens or somebody does something. Examples of economic events are:

- The Dow Jones fell.
- Toyota improved its hydrogen car.
- Prices are rising.
- National income went up.
- Consumers spend more.
- The government discusses income tax cuts.
- The central bank will raise the interest rate.

If we want to identify stories as the main element of narratives, the unit of analysis cannot be single words as in topic modeling, but should be events. A sentence that reports an event contains at least a subject (e.g. "the government") and a verbal predicate ("discusses") and often there is also an object ("income tax cuts"). Hence research methods for the analysis of narrative must be able to identify basic grammatical structures, which is not possible using bag-of-words models. It is already clear from this that the computerized analysis of narratives must be more sophisticated than topic modeling and requires more advanced methods of natural language processing.

The next step after the identification of events is to discover their temporal sequence. The smallest story consists one event occurring after another, e.g.

- The Dow Jones fell after the Fed had announced to raise the federal funds rate.
- Ben Bernanke had studied the Great Depression before he became a central banker.
- The financial system will collapse, if AIG fails.

We hence have to consider markers of time in the text, which can be different tenses, but also dates or temporal adverbs like "before", "after", "then", "yesterday" etc.

In a normal text, the story is not told in a single sentence, but in several sentences. Some sentences might provide details on the events or the agents and others might present evidence or examples. The following paragraph about Ben Bernanke taken from the Encyclopedia Britannica[xiv] illustrates this:

> *"He became a full professor in 1985 when he moved to Princeton University, and he served as a visiting professor at both New York University and MIT. Widely published on a range of economic issues—including macroeconomics, monetary policy, the Great Depression, and business cycles—Bernanke was awarded both a Guggenheim and a Sloan Fellowship, and in 2001 he became editor of the American Economic Review. The following year he was appointed to the Board of Governors of the Fed, and he became noted for thorough research and diplomacy when opinions among the governors differed. … In 2005 Bernanke was nominated by U.S. Pres. George W. Bush to succeed Alan Greenspan as chairman of the Fed. He took office on February 1, 2006. With his strong background in academia, Bernanke represented a clear break from previous Fed chairmen, who had usually come from Wall Street. While expected to uphold the style of fiscal management established by Greenspan, he brought certain important changes to the Fed, mainly in regard to inflation. Although his predecessor rejected inflation targeting, Bernanke preferred a stated inflation objective, which he believed would bring about economic growth and stability."*

The story can be summarized easily in one sentence: "Ben Bernanke was a respected academic before he became a central banker". In order to be a narrative, the story must have a specific meaning. We mentioned earlier that the fact that Bernanke had worked on the Great Depression as a scholar was interpreted to be of help during the financial crisis. Another narrative is presented in the paragraph from the Encyclopedia Britannica above. That text emphasizes the academic background because Bernanke broke with the policy style of his predecessor Alan Greenspan and replaced the fiscal management approach by inflation targeting. We could summarize this as narrative as follows:

"Ben Bernanke was a respected academic who had participated in the rules-vs.-discretion debate on monetary policy before he became a central banker. As a central banker, he replaced Alan Greenspan's `just do it approach´ by inflation targeting which is supported by many monetary theorists."

While this narrative may make a lot of sense to people who are familiar with the academic debate on monetary policy in the 1990s and 2000s, non-experts probably do not know what the `rules-vs.-discretion debate´ was about and what it has to do with Alan Greenspan. Depending on who uses this narrative, it can transport a critique of the `magician´ Greenspan and his claim of being a master of the `art of monetary policy´ as opposed to the `science of monetary policy´ in the subtext. Note that we use subtext ourselves here, when we talk about "the magician", "the art of monetary policy" and "the science of monetary policy", which are not mentioned in the narrative, but may be triggered by the term "just do it approach". Our approach is clearly hermeneutic here, but we argue this is necessary in order to capture the sensemaking element of narratives. The more a narrative is embedded into the collective belief system of a group, the fewer elements of that narrative must be explicated in a specific text to evoke said narrative (Meer, 2021). It is this hermeneutic nature of narrative sense making that poses the greatest challenge to the automated detection of narratives in texts.

Once a narrative with a certain meaning has been identified somehow, it might be easier again to analyze quantitatively how often it is used and by whom. It appears possible to characterize a specific narrative by typical phrases or terms and by the tone of the text. A researcher would have to identify the full narrative first and then define identifying keywords. It is necessary, however, also to measure whether the tone of the text is positive or negative, because both a supporter and a critic can use the same narrative. Instead of using topic modeling, topic classification approaches might be better suited to analyze the question, which groups share the same narratives. But given that topic classification methods are established, the identification of narrative-sharing groups should not pose fundamental problems.

Detecting the emergence of a shared narrative from the interaction of individual narratives, which are all slightly different, might be a bigger challenge. One way of doing this might be to show that distinct individual or sub-group narratives have a joint core and distinguishing features. Emergence might mean that the core of the narratives incorporates more and more new elements. Furthermore, the number of distinguishing features should decline over time. The main problem is to identify which narratives are similar, but not identical. Again, this might be more a hermeneutic task than a statistical one.

It can be difficult to say whether a text suggests a certain action. In some cases, the author of a text may give a direct recommendation to do something, e.g. buy a stock or vote for a politician. These cases are easy, but relatively rare. More frequently, financial analysts do not tell their readers directly to invest in a certain company, but say that the "outlook is good'' or that there are "opportunities". Similarly, a political commentator is unlikely to say "vote X", but might criticize politician Y and praise politician X. The reader nevertheless can interpret this as a suggestion to buy the stock or vote X, because he or she is looking for guidance from experts. The suggestion is often implied by the type of the text: it is the function of commentaries and opinion pieces to suggest a certain action, even if it is not literally expressed in the text. In many cases, however, the suggested action cannot be found directly in the words of a narrative, but, again, in the subtext and the context in which a narrative is used. In those cases, the perceived suggestion to perform a certain action is the product of the receiver's belief system. For example, the interpretation of the phrase "politician X wants to tax the rich more in order to finance public health care" depends on the context in which it is used and the belief system. For some readers, it may just be a factual statement without any suggestive character. However, if the phrase is used in a political context it can convey the message "vote X" or "do not vote X", depending on whether the writer and the reader believe that the described policy is good or bad. A way how researchers can deal with this ambiguity is to categorize the text by function, context and/or authorship and to assume that some texts contain suggestions while others do not.

## 5.2 Overcoming the identification problem

It is our position that, at this point, progress in the identification of economic narratives requires a mix of methods. In this section, we will introduce some important empirical methods and discuss their advantages and drawbacks with regards to identifying narratives. We will also highlight some important recent examples from the applied economics literature that successfully identify narrative building blocks through a mixed methods approach.

Topic modeling, despite being highly popular, can only be a first step in this empirical analysis because some elements of the syntactic structure underlying narratives elude topic models by construction. By using strictly document-level representations of words, topic models operate on a high level of linguistic abstraction. This should not be considered a "bug" or a weakness per se, since it is this reduction of complexity that allows topic models to extract economically relevant features from large collections of text. When topic modeling is applied to texts from the mass media or social media, we might be able to identify public discourses about economic topics and track their development over time. In these discourses, several narratives about the same topic might compete with each other.[xv] But in order to uncover these competing narratives in a discourse, topic modeling must be complemented by other methods from the toolkit of natural language processing (NLP). Those methods might fail to recover high-level textual elements (like public discourses) but could help in identifying finer narrative patterns in specific texts and sentences.

NLP offers a vast toolkit of well-established and highly specialized methods from computational linguistics that could complement topic modeling well.[xvi] Gentzkow et al. (2019) offer a review of such methods in an economic context. As the examples provided in chapter 5 show, economic narratives are often connected to institutions ("the government", "the central bank"), people ("Ben Bernanke") and economic concepts ("consumers", "prices"). When identifying narratives in a vast corpus of documents or a high-volume news feed, Named Entity Recognition (NER) models (Borthwick et al., 1998) can be used to pre-select interesting documents and categorize them according to the economic entities or concepts they pertain to. Benner et al. (2022) combine topic modeling with the named entity concept to track the discourse about some central economic agents over time. An economics-specific NER-model could even identify protagonists of an economic story in specific sentences. Since not all economic narratives necessarily refer to entities, syntactic parsing methods (Marcus et al., 1993; Jurafsky and Martin; 2020) can complement NER in uncovering subjects of economic discussions by labeling semantic roles. Part-of-speech-tagging (POS-tagging) is a more fundamental labeling method that classifies words according to their word class (Merialdo, 1994). Ash et al. (2021) use a combination of semantic role labeling and POS-tagging to identify narrative building blocks that follow the structure "who does what to whom?". While, from the standpoint of our definition, their method is not yet akin to identifying full narratives, it does detect events carried out by a certain group of entities and thus presents a promising methodological avenue.

Ash et al. (2021) also use clustered word embeddings to reduce the feature space of the narrative building blocks identified. Word embeddings are high dimensional vectors that capture syntactic as well as semantic relationships between words. They are based on the distributional hypothesis (Harris, 1954) from linguistics that postulates how a word derives its meaning through the specific context it is used in.[xvii] Word embeddings can be used to investigate word similarities and perform algebraic operations. A canonical calculation that illustrates the striking results that can be attained using word embeddings is the fact that $w2v^{king} - w2v^{man} + w2v^{women}$ results in an embedding that is closest to $w2v^{queen}$. Closeness is determined using the cosine of the angle between a pair of embedding vectors. However, it is important to consider that "meaning" is not an aggregate but a high-dimensional phenomenon. As such, it is unlikely that "similarity" between two words can be sufficiently captured by computing a single scalar. In an economic context, the limits of this oversimplified notion of similarity can be seen by the fact that $w2v^{increase}$ is among the most "similar" vectors to $w2v^{decrease}$. By virtue of the distributional hypothesis, this oddity is only logical, since "increase" and "decrease" will often be surrounded by the same context words. But it highlights the fact that such models cannot necessarily be leveraged for economic applications without any further research.

Methods like POS-tagging and NER are task specific in the sense that they can solve clearly defined linguistic problems. While these methods can extract well-defined features that a researcher deems important for narratives, they cannot by themselves extract full narratives or even elements of narratives as depicted in Figure 3. Ash et al. (2021) use a pipeline of several methods operating in tandem on a syntactic level to identify what we consider *events* of a story by extracting two entities that are connected through a verb. Their RELATIO-method offers promise because it allows for the identification of a latent core element of narratives (the *event*). It does not, however, tackle the important element of sense-making. As discussed in Section 3.2, sense-making is a highly complex cognitive and emotional process that lacks the kind of unambiguous, one-to-one textual counterpart one could hope to identify empirically. But the compression of information that is at the heart of narrative sense-making tends to involve the construction and assignment of causal relationships between events and entities. This causal connection removes ambiguity and, as such, constitutes much of the attraction that narratives offer to the human brain.

Lange et al. (2022) try to exploit this regularity by pre-filtering the paragraphs of text they use for explicit causal cues before applying a modified version of RELATIO. Restricting analyses to *explicit* causal cues, however, is not enough as it will surely yield high type II error rates. Causal relationships can by no means only be established explicitly. Consider the following example: "After Covid-19 started to spread throughout Wuhan, production in manufacturing came to a sudden halt in the city." Everybody who has experienced the onset of the Covid-crisis will clearly establish a one-way causal connection from the example above. This connection, however, is expressed by using a temporal conjunction. The recent powerful class of transformer language models may provide a solution for this problem. Models like BERT (Devlin et al., 2019) can be quite effective for various kinds of supervised text classification tasks because they provide a powerful general language model that represents words as context dependent embedding vectors. The model can then be fine-tuned for any number of classification tasks. Khetan et al. (2021) develop a version of BERT that is able to recognize both explicit and implicit causal structures. Through the selection of training examples with which fine-tuning is conducted, the researcher is also able to enrich and adapt BERTs vocabulary with regards to domain-specific language. This process may also help to quantify the social aspect of narratives by adapting BERT to a group-specific understanding of certain issues.

All of the aforementioned NLP-methods are vast fields in themselves and the papers mentioned herein should be considered examples from substantive literatures. It is clear that the tools to attain a richer representation of texts and move towards identifying narratives in addition to topics are available. Creative adaptations and combinations of these existing methods are already proving fruitful in extracting narrative elements in an economic context.

**Non-textual elements of sense-making**

Some components of economic narratives are unlikely to be identifiable with observational text data from news reports or social media. The interaction of narratives with human sense-making and belief systems on a psychological level can perhaps best be further understood using experimental designs. Antinyan et al. (2021) exploit the COVID-19 pandemic as a natural experiment to investigate how exposing individuals to the "lab accident" hypothesis or the "zoonosis" explanation influences their beliefs regarding the origin of the virus and loosely related political and socio-economic issues. They find that the "lab accident" prime in particular not only causes increasing adherence to the corresponding "lab accident" theory but also enforces negative views on foreign trade, climate change mitigation and science. Such spillover-effects indicate that narratives powerfully impact the belief system of individuals through their sense-making capabilities.

Fyshe et al. (2014) develop a vector space model similar to word2vec by incorporating brain activation data recorded through Functional Magnetic Resonance Imaging (fMRI) to attain a more complete representation of word semantics than would be attainable only through textual data. This rather extreme example highlights the fact that the sense-making aspect of language in general and economic narratives in particular transcends written language.

# 6 Wider implications of narratives in economics

In this section, we present some thoughts on what economists might take away for their future work from the host of literature of different origin. Our objective is not to advise other economists to do a certain kind of research or to adopt a certain view. We rather want to point to issues and questions that arise if "narrative" is defined as a concept in a profound way and not only used casually as a term with many vague meanings. Possible implications concern the discipline of economics itself, but also the exchange with other disciplines.

## 6.1 Implications for economics

Narratives are not only an object of economic study as suggested by Robert Shiller, but play a role in the way economists work, too. This insight results from the work of Deidre McClosky mentioned in the introduction. In contrast to "metaphor", McCloskey (1983) does not use the term narrative, when she talks about the use of rhetoric in economics, but narratives or stories are obvious rhetorical devices. She argues that *modernism* as the official methodology is neither possible nor adhered to by many economists in their actual practice. Instead of applying the so-called "Scientific Method" to find the truth, economists try to persuade others of their ideas. The possibility of falsification as part of the Scientific Method is very limited in economics. Therefore, economists cannot obtain certainty by proving what is right or wrong, but can only weigh reasons to arrive at conclusions that are more or less probable or plausible. This weighing of arguments is the actual meaning of rhetoric, in contrast to the common understanding of rhetoric as the use of verbal tricks or dishonest sophistry. McCloskey (1983) defines rhetoric as exploring thought by disciplined conversation. The most important rhetorical device in economics is the metaphor, i.e. the description of something by referring to something else that is considered to be similar. Economics is metaphorical, because it uses models that are claimed to be similar to reality to describe the economy. By making unrealistic, simplifying assumptions, economists create a kind of fiction with their models. Stories are needed to relate the abstract models to reality again.

According to McCloskey (1983, 1984), economics would benefit in several ways, if economists accepted the inevitableness of rhetoric in their discipline. First of all, it would remind economists that their metaphors are fiction, not reality. As argued before, fiction can be very helpful for learning by performing mental simulations. But deriving policy recommendations from models whose assumptions are not met, can be harmful to society. Quiggins (2012) discusses how economic ideas such as the Great Moderation, the Efficient Market Hypothesis, trickle-down economics or privatization and excessive financial market liberalization derived from unrealistic models[xviii] played a role in the Global Financial Crisis 2007 - 2009. Based on our concept of collective economic narratives, one could argue that the following narrative is a scientific narrative based on a fictional shared mental model of economists: "The financial sector is inefficient due to strangling regulation. Privatization and financial deregulation improve efficiency. A more efficient financial sector stimulates economic growth and improves societal welfare." This narrative was very strong both in economics and in policy circles especially during the 1980s and 1990s and contributed not only to the Global Financial Crisis, but to the smaller crises that occurred between 1980 and 2010. Another benefit could be that the quality of scientific debates might rise, if economists acknowledged that academic discourse and teaching are communicative processes with the aim of persuading others, for which a good and well-chosen rhetoric is helpful. Finally, McCloskey believes that the "foreign relations" to other disciplines might improve, if economists replaced the problematic modernist methodology by better rhetoric and literary thinking. This methodological reorientation would move economics away from the natural sciences and closer to the social sciences and the humanities which might be its proper place.

In a similar vein, Arjo Klamer (1988) stresses the importance of the rhetorical component in economic theorizing. He denies the notion of the science of economics being value-free simply because economists as people necessarily possess a value system that influences their ideas about the economy in some way. Klamer thus places great emphasis on dissecting the manner in which economists argue about the economy, how they talk about themselves and how they reflect on doing economics. That economics cannot be value-free was already discussed by Kenneth Boulding (1969), who argued that scientific communities share common value systems and that especially the social cannot simply observe an objectively given reality, but inevitably interact with their object of study. Even earlier,

Gunnar Myrdal criticized economic analysis for its hidden value content and its claim to be value-free (see Dykema 1986).

## 6.2 Interdisciplinary exchange

The relationship between economics and other disciplines is a difficult one. In general, economics is rather isolated from other social sciences, let alone from the humanities. If economics engages with other fields such as geography, sociology, political science or psychology, it often happens from a vantage point of superiority of the economists and is frequently perceived as "economic imperialism" by the other disciplines (Fourcade et al. 2015). Economic imperialism can be characterized by economists' attitude that other disciplines have interesting problems, but not the right methods to deal with them. Gary Becker was a pioneer in applying the economic approach, which he defined as "the combined assumptions of maximizing behavior, market equilibrium, and stable preferences, used relentlessly and unflinchingly" (Becker 1976, p. 5) to questions such as division of labor in the family, crime, discrimination or addiction. Apart from mathematical and statistical methods, economists are rather reluctant to import research methods from other disciplines.

Another wave of economic imperialism could happen again with narrative economics, if economists do not care about how other disciplines think about narratives. As our survey has shown, narrative economics has not spent much effort on learning what others know about narratives so far. The very starting point of our paper, namely that economists do not even define properly what they mean by narrative, is revealing in this context.

A scientific approach in economics with regards to narratives would benefit greatly from the insights in cognitive and affective science mentioned in Section 3.1 and 3.2. Narrating seems to be a fundamental mode of the human brain. The narrative brain suggests that humans are deeply social beings, which are not adequately described by the individualistic homo economicus that still is the base of most mainstream economic theorizing. If narrative economics wants to learn from cognitive and affective science, it must rethink its model of human behavior. First of all, humans should not be treated as isolated individuals that care only about themselves. There is a deep human interest in others and the desire not only to communicate information for selfish purposes, but also experiences and emotions. Second, the idea of fixed preferences is dubious if peoples' belief systems and even their brains interact with narratives. Finally, if cognitive and emotional processes interact, it is hard to maintain that fully rational optimization is the most representative assumption about human behavior. Of course, all of this is not new and by now well established in behavioral economics (see Cartwright, 2018). But the perfectly rational homo economicus is still the default model in economics and behavioral economics is often more a collection of deviations from the homo-economicus model than an alternative to it. The insights from cognitive narratology might provide a starting point for the development of alternative models of human behavior in economics.

Several authors appeal for a true dialogue between economics and the humanities, which might result in a better kind of economics called "humanomics" (McCloskey 2016, 2021, Morson and Schapiro 2017, Smith and Wilson 2019). Morson and Schapiro (2017) discuss what economics could learn from the humanities, without denying that economics has its merits and that the humanities could learn something from economics, too. They identify three areas in which the humanities could enrich economics: "with stories, a better understanding of the role of culture, and a healthy respect for ethics in all its complexity" (Morson and Schapiro 2017, p. 13). As discussed in the previous section, economists use stories to explain their models, but this is not what Morson and Schapiro have in mind. They argue that economics, including its behavioral branch, has a simplistic model of humans and that great novels could inform economists to create richer models of human behavior. In particular, economists might learn that humans are not only social beings as opposed to isolated individuals. They are also cultural beings whose identity and behavior are always contingent, embedded in a cultural context and history-dependent. Literary works stress particularity and the irreducibility of human experience to any single theory. Humans and their behavior cannot be explained from one perspective and simple models. By narration we can get an understanding of human complexity. Better economics hence requires narrativeness. Finally, economists can benefit from the humanities whenever they make policy recommendations, because those always involve complex ethical questions. Morson and Schapiro argue that ethical questions cannot be reduced to any theory or a set of simple rules, but require good judgment and wisdom. Good judgment follows from case-based reasoning that takes the particularities of the situation into account. They hence recommend a literary and novelistic approach to ethical questions that could improve economic policy recommendations.

# 7 Conclusions

While a full narrative turn has not occurred yet in economics, the interest in narratives, their origins, consequences and dynamics has increased considerably in the profession as we have shown in our literature review.

So far, most progress was made on the methodological side, in particular in the application of NLP methods and the formal modeling of narrative dynamics. But the full potential of narrative economics – which is gaining a new perspective on the economy and on how to do economics – is still to be realized. We contribute to this next step by offering a precise definition of collective economic narratives, which fills a gap in the existing literature. Our definition of collective economic narratives as sense-making stories shared by members of a group, which emerge in social interaction and suggest actions, is rooted in complexity economics, but also in a variety of other fields, and adds three features to existing definitions. First, we are explicit about what we mean by "story". For us, a story is the report of a temporal sequence of events and not just a collection of terms that are somehow related. Second, we explain precisely what we mean by "sense-making" and argue that sense-making happens on the basis of people's belief systems which consist of their mental models of how the world functions and their value systems. Hence our approach emphasizes the inherent subjectivity of narratives, which has important epistemological and ontological implications. On the epistemological side, we have to deal with the question of what researchers can know about the sense-making process and the belief system of agents and how they can acquire such knowledge. It seems inevitable to allow for some hermeneutics in the study of narratives. Ontologically, the subjective dimension of narratives means that there is not just one, objective economy. If it is correct that economic behavior depends on narratives, which, in turn, depend on the subjective belief systems of economic agents, then economic decisions are made on the basis of how agents perceive the economy and not on how it actually is (whatever this means in this case). This leads to our third insight, which is that narratives emerge and evolve over time. At any given time, there is a plethora of individual narratives about all kinds of economic issues as people try to make sense of what they observe. Social processes lead to the emergence of collective narratives which coordinate the behavior of groups and can have the power to drive the economy at the aggregate level. While it can happen that a single narrative is dominant for a while, it might be the more frequent case that multiple narratives compete and interact with each other. The emergence of collective narratives is both a cause and a consequence of the permanent evolution of the economy.

In order to uncover these competing narratives in a discourse, economists must tap into the recent – and probably many forthcoming – advances in empirical language modeling. Research in narrative economics can benefit from importing these methods but also from co-developing new and specific tools in exchange with computer scientists and computational linguists.

Narrative is not "just" a type of text, but a universal tool of the mind that drives social dynamics in a fundamental way. Therefore, economists can also benefit and learn from cognitive and social psychologists as well as sociologists on how to approach individual and collective sense-making. In order to reap its full potential, narrative economics needs interdisciplinary exchange.

## Acknowledgements

We are grateful for stimulating discussions and feedback from Henrik Müller, Dorothee Meer, Sebastian Susteck, Elliott Ash, Philine Widmer, Germain Gauthier, Tobias Schmidt, Jonas Rieger, Niklas Benner, Lena Marie Hufnagel and Nico Hornig. We also wish to thank three anonymous referees for their valuable comments, which extended our perspective considerably. Of course, we are to blame for any remaining weaknesses.

## Notes

[i] Denzau and North (1994) refer to W. Brian Arthur and complexity concepts such as path dependence, adaptation or punctuated equilibrium. Dolfsma et al. (2011) discuss the role of communication in Niklas Luhmann's theory of social systems (Luhmann 1995), which is also related to complex systems.

[ii] It is, of course, also possible to define and conceptualize economic narratives from a rational choice perspective. Schwartzstein and Sunderam (2021) present a framework where "model persuaders" influence receivers' beliefs by proposing data-organizing models. The receivers perform a Bayesian hypothesis test for their choice of the proposed models. Narratives could be interpreted as the persuaders' models. Schiemann (2007) argues that approaches informed by cognitive psychology and behavioral economics provide a more accurate account to analytic narratives.

[iii] The literature on quantitative text analysis calls such unstructured collections of terms *topics.* Section 5 features a discussion on the appropriate use of topic modeling methods in the context of narrative economics*.* However, the term *topic* has different definitions in different fields. In linguistics, a topic is the phrase in a clause that the rest of the clause is about. In discourse analysis, the topic is what a discourse is about.

[iv] A word of caution must be noted on the distinction between cognition and emotion: While the two concepts have traditionally been viewed as independent entities in the cognitive sciences, more recent behavioral and neuroscientific research calls this paradigm into question. It instead suggests a more complex, interdependent relationship between cognition and emotion (Storbek and Clore 2007, Duncan and Barrett 2007).

[v] https://www.ft.com/content/a3a2542f-0feb-4596-8357-30ab965697d6

[vi] Note that the following sentences are potential names for the narratives, not the narratives themselves. We formulate the narratives in the text.

[vii] https://www.newyorker.com/news/steve-coll/lessons-of-the-great-depression , https://nymag.com/news/businessfinance/bottomline/57177/

[viii] https://www.federalreserve.gov/newsevents/speech/bernanke20100408a.htm

[ix] https://www.ineteconomics.org/perspectives/blog/macroeconomics-predicted-the-wrong-crisis, https://www.forbes.com/2009/02/18/depression-financial-crisis-capitalism-opinions-columnists_recession_stimulus.html?sh=31a3653722ef,

[x] https://www.economist.com/leaders/2008/10/16/capitalism-at-bay, https://www.democracynow.org/2008/10/6/naomi_klein

[xi] https://www.newyorker.com/magazine/2008/11/17/the-new-liberalism

[xii] https://www.wsws.org/en/articles/2008/09/lehm-s16.html

[xiii] https://www.theguardian.com/business/blog/2010/sep/02/ben-bernanke-financial-crisis-inquiry

[xiv] https://www.britannica.com/biography/Ben-Bernanke

[xv] Müller et al. (2018) refer to topics identified by LDA in corpuses of news articles as "mean media narratives", a testament to the fact that such topics will generally contain not one but a collection of narratives, as defined above. Those narratives will be similar with regards to subject matter but may reflect different, even opposing viewpoints and value judgements.

[xvi] NLP is a rapidly moving field and state-of-the-art generative NLP-models are among the most complex in deep learning today even though those models mostly produce black boxes that predict well but explain poorly. OpenAI´s GPT-3 language model uses 175 billion parameters (see Floridi and Chiriatti (2020) for a discussion), which is further testament to the level of complexity that is inherent in language.

[xvii] A full discussion of language models is beyond the scope of this paper. For an up-to-date review on language models and their uses in economics, we refer to Ash and Hansen (2022) and the sources cited therein.

[xviii] Note that the critique lacking realism especially in macroeconomic modeling is shared by many eminent economists like the Nobel laureates Wassily Leontief and Paul Romer (see Leontief 1971 and Romer 2016).